\documentclass[twocolumn,aps,prl,floatfix,superscriptaddress,10pt]{revtex4-2}

\usepackage{amsmath,amssymb,amsfonts,graphicx,braket,hyperref}
\hypersetup{colorlinks=true,citecolor=magenta,linkcolor=red}

\graphicspath{{./},{./figs/}}

\usepackage{graphicx} 
\usepackage[caption=false]{subfig}
\usepackage{physics}
\usepackage{bbold}
\usepackage{comment}
\usepackage{booktabs}
\usepackage{xcolor}

\usepackage[normalem]{ulem}

\begin{document}

\title{Topological phase transitions in a constrained two-qubit quantum control landscape}

\author{Nicol\`o Beato}
\email{nbeato@pks.mpg.de}
\affiliation{Max Planck Institute for the Physics of Complex Systems, N\"othnitzer Str.~38, 01187 Dresden, Germany}
\author{Pranay Patil}
\email{pranay.patil@iitm.ac.in}
\affiliation{Theory of Quantum Matter Unit, Okinawa Institute of Science and
Technology Graduate University, Onna-son, Okinawa 904-0412, Japan}
\affiliation{Department of Physics, Indian Institute of Technology Madras, Chennai 600036, India}
\author{Marin Bukov}
\email{mgbukov@pks.mpg.de}
\affiliation{Max Planck Institute for the Physics of Complex Systems, N\"othnitzer Str.~38, 01187 Dresden, Germany}

\date{\today}

\begin{abstract}
In optimal quantum control, control landscape phase transitions (CLPTs) indicate sharp changes occurring in the set of optimal protocols, as a physical model parameter is varied.
Here, we demonstrate the existence of a new class of CLPTs, associated with changes in the topological properties of the optimal level set in a two-qubit state-preparation problem.
In particular, the distance distribution of control protocols sampled through stochastic homotopic dynamics reveals discontinuous changes in the number of connected components in the optimal level set, as a function of the protocol duration.
We demonstrate how topological CLPTs can be detected in modern-day experiments.
\end{abstract}

\maketitle

\begin{figure}[t!]
\centering
\includegraphics[width=.48\textwidth]{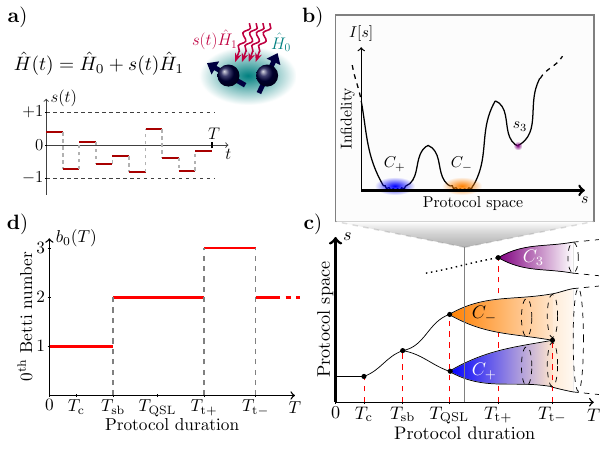}
\caption{
Topological transitions in the two-qubit quantum control landscape in Eq.~\eqref{eq:H}. 
\textbf{a)} Piecewise-constant protocols $s(t)$ control a system $H_0$ by modulating the external drive $H_1$ for a total duration $T$ (inset panel (b)).
\textbf{b)} Illustration of the optimization landscape $I_T[s]$ for a fixed protocol duration $T$.
Depending on $T$, the landscape possesses a different number of basins of attraction; in this schematic, two basins are associated with the connected components $C_+,C_-$ of the infinite-dimensional optimal level set and one with the local trap $s_3$.
\textbf{c)} Sketch of the optimal level set $\{\mathrm{arg\,min}_sI_T[s]\}$ as a function of $T$. Topological transitions at $T_\text{sb}$, $T_\text{t+}$, and $T_\text{t--}$ mark sudden change in the number of connected components $b_0(T)$.
\textbf{d)} Topological CLPTs correspond to jump-discontinuities of a topological invariant known as the $0^\text{th}$ Betti number $b_0(T)$ as a function of $T$; the $T_\text{QSL}$ transition does not affect $b_0(T)$.
}
\label{fig:intro}
\end{figure}

Optimal control theory is central for modern-day quantum technologies where quantum properties are preserved and exploited through precise and efficient hardware manipulations \cite{
koch2022quantum,
acin2017roadmap,
dalessandro2021introduction,
boscain2021introduction}.
Despite the numerous advances in recent years, complex quantum systems remain difficult to control.
The procedure aiming at finding the optimal control protocol (Fig.~\ref{fig:intro}a), often terminates in a sub-optimal solution, while the true-optimum remains inaccessible or extremely difficult to find \cite{
ge2022optimization,
cunningham2024investigating,
Bukov19_Glassy}.
This situation is widespread in optimization problems \cite{
gamarnik2022disordered,
altieri2021properties,
zdeborova2016statistical,
krzakala2007landscape,
stein1995broken,
montanari2024,
fyodorov2022,
urbani2023,
urbani2024} 
where one is typically interested in characterizing the problem's complexity.
Common approaches analyze the critical sets of the optimization landscape -- stationary points of the cost function that assigns a figure of merit to each control protocol (Fig.~\ref{fig:intro}b)
\cite{
ge2021,
ros2023high,
rabitz2004quantum,
kent-dobias2023how,
kent-dobias2025,
mezard2002analytic}.
Among the critical sets, one distinguishes the optimal level set, consisting of protocols with the best performance in the control task.

The quantum control landscape depends on external model parameters, e.g., the total protocol duration. Varying these can lead to sharp deformations of the optimal level set; in this case, specific order parameters designed to capture statistical properties of optimal protocols \cite{
Bukov18_Reinforcement,
hu2012stability,
Nishimori01,
mezard1987spin},
exhibit non-analytic points (cusps, jump-discontinuities, etc.) called ``Control Landscape Phase Transitions" (CLPTs) \cite{Bukov18_Broken}, in analogy to transitions in statistical mechanics.
As the optimization complexity depends on the structure of the landscape, a thorough understanding of CLPTs is of central importance for the characterization of quantum control problems.
So far, CLPTs have been associated with local changes in the optimal level set, i.e., changes involving the neighborhood of isolated optimal protocols \cite{beato2025toward}. In this work, we raise the question of whether CLPTs can exist, associated with changes in the topology (i.e., global changes) of the optimal level set.

Investigating topological properties requires probing the global structure of the systems of interest 
\cite{
cong2024enhancing,
copenhagen2021topological,
bridgeman2016detecting,
jiang2012identifying,
levin2006detecting}.
In the case of quantum control landscapes, an analytical parametrization of the optimal level set is generally missing, so methods of Differential Topology are not applicable~\cite{
dalessandro2021introduction,
SchaettlerLedzewicz12}.
Conversely, numerical methods do not necessarily give information about topological properties; e.g., ``bang-bang" protocols (alternating sequences of pulses of fixed amplitude) do not allow for generic continuous transformations (homotopies), used in a topological analysis.

In this work, we study a prototypical quantum control problem, and report a new class of control landscape phase transitions, characterized by changes in the topological structure of the optimal level set (Fig.~\ref{fig:intro}c).
We develop methods to explore different connected components via stochastic homotopic transformations, and find discontinuous changes in their number -- a topological invariant known as the $0^\text{th}$ Betti number (Fig.~\ref{fig:intro}d) \cite{hatcher2005algebraic}.
Finally, in the End Matter we discuss how modern-day quantum experiments can track topological CLPTs via standard measurements of observables (e.g., qubit polarization).

\textit{Model.---}Consider a state preparation problem associated with the two-qubit Hamiltonian
\begin{equation}
    \hat H(t) = -J \hat S_1^z \hat S_2^z -h_z (\hat S_1^z + \hat S_2^z) - s(t) \,h_x(\hat S_1^x + \hat S_2^x),
    \label{eq:H}
\end{equation}
where $\hat S^\alpha (\alpha{=}x,y,z)$ are the $\mathfrak{su}$(2) generators describing the qubit, and $s(t)$ is the control protocol. Let us fix $J{=}2,h_z{=}1,h_x{=}\sqrt5$ and set
the initial $\ket{\psi_0}$ and target $\ket{\psi_*}$ states as ground states of $\hat H[h_x{=}{-}2]$ and $\hat H[h_x{=}{+}2]$, respectively. We restrict the protocol space to bounded functions $\abs{s(t)}{\le}1$, reflecting physical limitations in the output power of the applied external field \cite{Bukov19_Geometric}.

The infidelity $I_T[s] {=} 1{-}\abs{\bra{\psi_*}\hat U[s](T) \ket{\psi_0}}^2$ defines the cost function of the state-preparation problem; it depends on the duration of the quantum evolution $T$ ($t{\in}[0,T]$) and on the time-dependent protocol $s(t)$. The mapping $s{\mapsto}I_T[s]$ from protocol space to the interval $[0,1]$ defines the control landscape.  
This control problem possesses two characteristic durations of the quantum evolution, $T_c$ and $T_\text{QSL}$, associated with different types of instabilities leading to control phase transitions in the optimal level set $\{\mathrm{arg\,min}_sI_T[s]\}$ \cite{beato2025toward}.
At $T_c$, the optimal control protocol becomes singular \cite{dalessandro2021introduction,boscain2021introduction,beato2025toward}.
$T_\text{QSL}$ is the critical duration, past which one can prepare the target with arbitrarily small infidelity via numerical optimization (QSL: quantum speed limit).
The specific choice of parameters $h_z,h_x,J$ affects the precise location of these CLPTs but does not lead to a qualitative change in the control phase diagram. 
In addition, this control problem possesses a $\mathbb Z_2$ symmetry-breaking transition: at $T{=}T_\text{sb}$ optimal protocols break the control symmetry $s(t){\to}{-}s(T{-}t)$; this CLPT depends on the choice of initial and target states $\ket{\psi_{0,*}}$, respecting the cost function symmetry $I_T[s(t)]{=}I_T[{-}s(T{-}t)]$ \cite{Bukov18_Broken}.

\textit{Methods.---}%
Control phase transition are a property of the control landscape and do not depend on the specific algorithm used to probe the landscape. 
We now generalize this concept to include topological CLPTs, which occur as discontinuous changes in the topological invariants of the optimal level set $\{\mathrm{arg\,min}_sI_T[s]\}$ (Fig.~\ref{fig:intro}c). 
Topological invariants capture global properties of topological spaces. For this reason, they are in general inaccessible by analysis limited to the local properties.

In this work, we focus on changes in the number of connected components $b_0{=}b_0(T){\in}\mathbb{N}$ (so-called $0^\text{th}$ Betti number).
To detect topological transitions, we use the Langevin-Monte Carlo algorithm (LMC)~\cite{beato2025toward}, which we now briefly recap. 
First, the time interval $[0,T]$ is divided in $L$ equidistant steps $\{t_1,\dots,t_L\}$ of length $\delta t {=} T/L$  (Fig.~\ref{fig:intro}a). LMC explores the control landscape using purely diffusive stochastic dynamics which updates the piecewise-constant protocol $s {=} (s(t_1),\dots,s(t_L))$. The optimization dynamics depends on two parameters: $\sigma$ sets the typical length of the local update $s_{n+1}(t) {=} s_n(t) {+} \xi(t)$ ($\xi(t)$ is a Gaussian-distributed random variable, with zero mean and variance $\sigma^2$); $\beta$ controls the Monte Carlo acceptance probability $e^{-\beta \Delta I_{n+1}}$ which depends on the infidelity increment $\Delta I_{n+1} {=} I(s_{n+1}) {-} I(s_{n})$ following the update.
We then run $R$ independent LMC simulations, starting from different initial protocols. We collect $R$ independent sets, $\{\mathcal S_\alpha\}_{\alpha=1}^R$, each containing $M$ sampled protocols: $\mathcal S_\alpha{=}\{s_1^{\alpha},\dots,s_M^{\alpha}\}$. Between two sampled protocols, we wait for $\Delta n {=} 2^{14}$ LMC iterations, corresponding to $\Delta n L$ attempted local updates
\cite{beato2025toward}.
As $\beta$ is increased, the LMC dynamics approximates a continuous deformation (i.e., a homotopy) of the protocol $s$ within the optimal level set. In the limit $\beta{\to}\infty$, each LMC run is strictly confined to a single connected component after waiting for sufficiently many iterations. We extract global information about the optimal level set by comparing properties of the independent sets $\{\mathcal S_\alpha\}$.

\begin{figure}[t!]
\centering
\includegraphics[width=.48\textwidth]{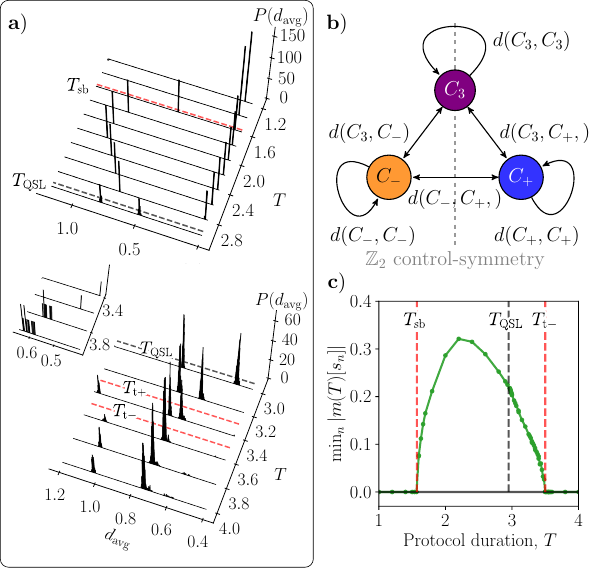}
\caption{
Analysis of the connected components of the optimal level set $\{\mathrm{arg\,min}_sI_T[s]\}$, before and after the quantum speed limit $T_\text{QSL}$ (black dashed line).
\textbf{a)} Density distribution $P(d_\text{avg})$ of the \emph{average} distance $d_\text{avg}$ in Eq.~\eqref{eq:d_a}, for different protocol durations $T$. Changes in the number of peaks in $P(d_\text{avg})$ mark topological transitions (red dashed lines).
\textbf{b)}
Graph summarizing structure of optimal level set for $T_\text{t+}{\le}T{\le}T_\text{t--}$. Connected components $C_+,C_-,C_3$ are represented by nodes while edges represent the relative distances $d(C_i,C_j),i{=}+,-,3$. A peak in the distance distribution $P(d_\text{avg})$ in (a) corresponds to at least one edge in the graph.
\textbf{c)}
The order parameter $\min_n \abs{m(T)[s_n]}$ in Eq.~\eqref{eq:m} identifies with high precision the location $T_\text{t--}$, associated with the merging $C_+ C_-{\to}C_\pm$.
LMC parameters: $L{=}64,\beta{=}10^6,\sigma{=}10^{-2}$.
}
\label{fig:fig2}
\end{figure}

First, we study the distance distribution between pairs of sets in $\{\mathcal S_\alpha\}$.
The squared $L^2$-distance between two protocols $s_1,s_2{:}[0,T]{\to}[{-}1,1]$ is defined as $d^2(s_1,s_2){=}T^{-1}\int_0^T\dd t\,(s_1(t)-s_2(t))^2$.
The \emph{average} distance $d_\text{avg}(\mathcal S_\alpha, \mathcal S_\beta)$ between two sets $\mathcal S_\alpha,\mathcal S_\beta$ is defined as
\begin{align}
    d_\text{avg}(\mathcal S_\alpha, \mathcal S_\beta) &=
    \frac{1}{M^2} \sum_{i,j=1}^M d(s^\alpha_i, s^\beta_j).
    \label{eq:d_a}
\end{align}
The distance $d(s_1,s_2)$ is a central concept in Topological Data Analysis \cite{carlsson2021topological,wasserman2016topological}, closely related to the protocol \emph{overlap}, $s_1{\cdot}s_2{=}T^{-1}\int_0^T\dd t\,s_1(t)s_2(t)$, routinely used to investigate spin glass physics \footnote{
In our case, $s_1,s_2$ are not normalized to unity nor restricted to $\{{\pm}1\}$ (as in the spherical and Ising model).
}; below, we leverage its properties to detect topological CLTPs.
For alternative definitions of the distance between the sets $\{\mathcal S_\alpha\}$ see \cite{Supplemental}.

From $\{\mathcal S_\alpha\}$, we obtain a total of $R(R{-}1)/2$ values, representing the distance $d_\text{avg}$ between the pairs of sets. The information about the different connected components $\{C_1,C_2,\dots\}$ of the optimal level set is contained in the distribution $P(d_\text{avg})$ (Fig.~\ref{fig:fig2}a). Notice that the distribution is a function of the protocol duration $T$; as $T$ increases, the optimal level set deforms, and the number of connected components changes: $b_0(T)$ exhibits discontinuous jumps (Fig.~\ref{fig:intro}c,d).

\textit{Topological Control Landscape Phase Transitions.---}%
The distribution $P(d_\text{avg})$, shown in Fig.~\ref{fig:fig2}a for $T{\in}[0,4]$, depends on the LMC parameters used to sample the sets $\{\mathcal S_\alpha\}$.
Here, we are interested in the features exhibited by $P(d_\text{avg})$ surviving the limit $M{\to}\infty$, $\beta{\to}\infty$, in this order \footnote{ 
The order of the two limits is important to guarantee that the distribution of protocols from the region around the optimal level set is converged, at fixed $\beta$.
}.
Each peak in  $P(d_\text{avg})$ sharpens into a delta function for $M{\to}\infty$, as the distance $d_\text{avg}$ between each pair $\{C_i,C_j\}$ is unique.
Hence, in the case of a finite number of samples, we identify the location of each peak with the average distance $d_\text{avg}(C_i,C_j)$ between two connected components $C_i,C_j$ (possibly $i{=}j$).
For $T{\in}[0,T_\text{sb}]$, $P(d_\text{avg})$ has a single peak located at $d_\text{avg}{\approx}0$ so that the optimal level set consists of a single connected component, which we call $C_0$; $C_0$ consists of an isolated protocol since the peak moves to $d_\text{avg}{=}0$ as $\beta{\to}\infty$ \cite{Supplemental}.

This behavior changes past $T_\text{sb}$, where $P(d_\text{avg})$ acquires a second peak centered at $d_\text{avg}{\approx}0.6$ for $T{=}1.6$ (cf.~Fig.~\ref{fig:fig2}): the optimal level set undergoes a topological transition and now possesses two distinct connected components, called $C_+,C_-$.
The location of the first peak converges to $d_\text{avg}{=}0$ as $\beta$ is increased \cite{Supplemental}: the two connected components are two isolated optimal protocols, called $s_+,s_-$.
This bifurcation is associated with the breaking of the control symmetry, i.e., $s_\pm(t){\ne}{-}s_\pm(T{-}t)$ \cite{Bukov18_Broken,li2023optimal}.
Here, the topological nature of the $T_\text{sb}$ CLPT is apparent: the number of connected components in the optimal level set jumps from one to two. 
Coincidentally, at the symmetry-breaking transition, the optimal level set consists of a single control protocol, so that the local geometry provides information about a topological change. 
More generally, the identification of topological transitions requires the study of non-local properties of the optimal level set.

The optimal level set changes dimensionality for $T{\ge}T_\text{QSL}$~\cite{beato2025toward}. Two connected components, containing an infinite number of optimal protocols, originate from $s_+,s_-$ (Fig.~\ref{fig:fig2}a): the peak located in $d_\text{avg}{\approx}0$ for $T{<}T_\text{QSL}$, suddenly moves to the finite value $d_\text{avg}{\approx}0.5$ for $T{>}T_\text{QSL}$ and its location is unaffected by increasing $\beta$.

Interestingly, the peaks corresponding to $d_\text{avg}(C_+,C_-)$ and $d_\text{avg}(C_+,C_+){=}d_\text{avg}(C_-,C_-)$ are well-separated for $T_\text{QSL}{\le}T{\le}3.4$ but indistinguishable for $T{\ge}3.6$.
We deduce that the components $C_+,C_-$ merge into a single connected component, called $C_\pm$, between $T{=}3.4$ and $T{=}3.6$. 
We define $T_\text{t--}{\in}(3.4,3.6)$ as the critical duration associated with the merging transition $C_+ C_-{\to}C_\pm$. This marks a new kind of CLPT, involving topological properties of the optimal level set, where the number of connected components $b_0(T)$ decreases by one unit \footnote{
    If the point(s) of contact between the two connected components at $T{=}T_{\text{t}-}$ is known, an analysis of the local geometry at this point(s) should also reveal the merging topological transition.
    However, the identification of the minimum cost pathway(s) between two connected components requires a non-local analysis; typically used algorithms are the \emph{string method} \cite{weinan2002} or the \emph{nudged elastic band method} \cite{henkelman2000climbing,henkelman2000improved}.
}.

Rather unexpectedly, for $T{\ge}3.2$ we observe two extra peaks appearing in $P(d_\text{avg})$: the first one is well visible at $d_\text{avg}{\approx}1.2$ while the second one is relatively smaller and located at $d_\text{avg}{\approx}0.4$
\footnote{
The different heights of the peaks are due to the different number of LMC runs confined in each connected component. In this case, LMC runs that sample the component $C_3$ are fewer in number than the ones exploring $C_+$ or $C_-$.
}.
We associate these additional peaks with the appearance of a third connected component in the optimal level set, called $C_3$.
Hence, we find a second type of topological CLPT at $T_\text{t+}{\in}(3.2,3.4)$, associated with incrementing the number of connected components in the optimal level set.
In \cite{Supplemental} we demonstrate that the third connected component $C_3$ originates from an isolated sub-optimal protocol $s_3(t)$ (cf.~Fig.~\ref{fig:intro}); $s_3$ is sub-optimal for $T{<}T_\text{t+}$ but turns into a global minimum at $T{=}T_\text{t+}$, consistent with Fig.~\ref{fig:fig2}a.

Even though one might expect a total of six peaks marking the average distances between the three connected components $\{C_+,C_-,C_3\}$, the $s(t){\leftrightarrow}{-}s(T{-}t)$ control symmetry reduces the number of distinct peaks in $P(d_\text{avg})$. 
In Fig.~\ref{fig:fig2}c we visualize the three connected components $C_+,C_-,C_3$ as nodes, and their relative distances $d_\text{avg}(C_i,C_j)$ as edges, $i,j{\in}\{+,-,3\}$. The control symmetry transforms $C_+{\leftrightarrow}C_-$ and $C_3{\leftrightarrow}C_3$
\footnote{
Notice $d(C_3,C_-) {=} d(C_3,C_+)$ since $C_3$ is left invariant by the $s(t){\leftrightarrow}-s(T-t)$ symmetry.
}, and hence we find the relations $d_\text{avg}(C_+,C_+){=}d_\text{avg}(C_-,C_-)$, $d_\text{avg}(C_+,C_3){=}d_\text{avg}(C_-,C_3)$ identifying two among the six relative distances.

In Ref.~\cite{Bukov18_Broken} the symmetry-breaking transition occurring at $T{=}T_\text{sb}$ has been related to a cusp in the protocol ``magnetization'' order parameter $m(T)[s]{=}T^{-1}\int_0^T\dd t\, s(t)$.
Now, we adapt this observable to investigate the topological phase transition at $T_\text{t--}$: the $s(t) {\leftrightarrow} {-}s(T{-}t)$ symmetry of the optimal protocols, enforcing a vanishing $m(T)[s]$, is broken for $T{>}T_\text{sb}$ and can be used to classify the two connected components $C_+,C_-$. 
We already know from the distribution $P(d_\text{avg})$ that the two symmetry-breaking components of the optimal level set merge somewhere in $T{\in}[3.4,3.6]$; therefore, we anticipate the same merging transition to be visible in the magnetization order parameter, as we now show. 

Past the $T_\text{QSL}$ transition, the magnetization associated with the protocol $s_n$ evolved by LMC, fluctuates in time.
This fact motivates us to consider the \emph{minimum} absolute magnetization registered during each LMC run 
\begin{equation}    
    \min_n \abs{m(T)[s_n]} = \min_n{ \abs{\frac1T\int_0^T\dd t\, s_n(t)} }.
    \label{eq:m}
\end{equation}
To restrict the study to the two connected components generated by the $T_\text{sb}$ transition, we exclude the LMC runs sampled from the third disconnected component $C_3$ present for $T{\ge}3.4$
\footnote{
This can be done, e.g., by excluding LMC runs corresponding to the $C_3$ peak in the distance distribution $P(d_\text{avg})$ (cf.~Fig.~\ref{fig:fig2}a).
}.

In Fig.~\ref{fig:fig2}c we see that the minimum magnetization achieved during each LMC run is non-vanishing only in the interval $[T_\text{sb},T_\text{t--}]$.
Hence, we can identify with high precision $T_\text{t--}{\approx}3.5$ as the critical duration at which $\min_n \abs{m(T)[s_n]}$ vanishes.
We confirm that, for $T{>}T_\text{t--}$, the two (previously distinct) connected components $C_+,C_-$ have merged and formed the single $C_\pm$; LMC can move between the positively and negatively magnetized regions of $C_\pm$ via homotopies while preserving vanishing infidelity 
\footnote{
Rigorously, the positively and negatively magnetized regions of $C_\pm$ are defined by $\{s\in C_\pm : m(T)[s]>0\}$ and $\{s\in C_\pm : m(T)[s]<0\}$, respectively.
}.

The results presented are obtained at a fixed number of protocol time steps, $L{=}64$. We do not observe qualitative changes in the results with varying $L$~\cite{Supplemental}.

\textit{Outlook.---}%
We have presented numerical evidence of a new class of control landscape phase transitions (CLPTs), involving topological properties of the optimal level set and independent of the symmetries(s) of the control problem. We have illustrated the underlying mechanism using a constrained two-qubit state-preparation problem.
In particular, we infer global information about the topology of the optimal level set by sampling optimal protocols via stochastic homotopic dynamics and considering (i) the distances between sampled sets, and (ii) the control symmetry-breaking order parameter.

The techniques and concepts we have developed are general and can be applied straightforwardly to detect CLPTs in generic control problems.
Our novel approach shows that, for protocol durations greater than the quantum speed limit, the optimal level set consists of a few connected components, each containing infinitely many homotopically equivalent optimal protocols. 
By tracking the number of connected components as the duration of the quantum evolution is changed, we discover two types of topological CLPTs: 
in the first type, distinct connected components $C_i,C_j$ undergo a merging process $C_i C_j {\to} C_{i+j}$; in the second type, a new connected component originates from a sub-optimal protocol (local trap) reaching global optimality. 

Our study raises the question about the existence of topological CLPTs involving \emph{higher} Betti numbers of the optimal level set; these may be detected, e.g., using more advanced techniques from Topological Data Analysis \cite{carlsson2021topological,wasserman2016topological}. 
For instance, we expect a change in higher Betti numbers to occur at the quantum speed limit transition, as a previous work suggests a drastic change in the shape of the optimal level set, from a single point to a high-dimensional truncated hyperboloid~\cite{beato2025toward}.

In the End Matter, using the reduced density matrix, we show that topological CLPTs are associated with homotopically inequivalent trajectories in Hilbert space. Hence, modern-day quantum platforms capable of simulating isolated two-qubit systems can detect topological CLPTs by measuring the qubit magnetization or entanglement entropy.

Our work marks a step toward an improved characterization of two-qubit quantum control landscapes as we show that the optimal level set consists of a small number of connected components.
Each connected component is associated with an average protocol, while fluctuations measure the size of the connected component.
Whereas we focused on a two-qubit control problem, our study reveals behaviors of optimal level sets that are likely generic. 

\textit{Acknowledgments.---}
Funded by the European Union (ERC, QuSimCtrl, 101113633). Views and opinions expressed are however those of the authors only and do not necessarily reflect those of the European Union or the European Research Council Executive Agency. Neither the European Union nor the granting authority can be held responsible for them.
Numerical simulations were performed on the MPIPKS HPC cluster.

\textit{Data availability.---}
The data and code associated with this manuscript are available under \cite{data_availability}.

\bibliography{bibliography}

\pagebreak
\clearpage

\begin{widetext}
\begin{center}
\textbf{End Matter}
\end{center}
\end{widetext}

\textit{Observing Topological Control Phase Transitions in Quantum Experiments.---}
Topological CLPTs give rise to distinct signatures, readily observable in modern-day experiments. 
We can directly visualize the topological transitions by considering the single-qubit reduced density matrix $\rho_1(t) = \Tr_2[\rho(t)]$, where $\rho(t){=}\dyad{\psi(t)}$ and $\Tr_2$ is the partial trace over the second qubit 
\footnote{
Due to the qubit exchange symmetry of the Hamiltonian, tracing out the second or first qubit yields the same reduced density matrix.
}.
Because the reduced density matrix contains information about the system's state, we expect topological CLPTs to affect expectation values of single-particle observables 
\footnote{
In general, the choice of the quantum observable(s) for the detection of topological CLPTs depends on the
trajectories followed by the system along optima in different connected components.
The fact that the single-qubit reduced density matrix contains partial information about the system’s state is not limited to two-qubit systems. 
For this reason, we expect topological CLPTs to affect expectation values of single-particle observables even for controlled systems with many interacting qubits.
Nevertheless, if this is not the case, quantum observables involving more qubits are required.
}.
Here, we write $\rho_1(t){=}  \mathbb I/2 {+} \vec n_1 \cdot \hat {\vec{S}}_1$, and plot the trajectory of $\vec n_1(t)$ inside the Bloch sphere, $\abs{n_1}{\le}1$.

\begin{figure}[t!]
\centering
\includegraphics[width=.235\textwidth]{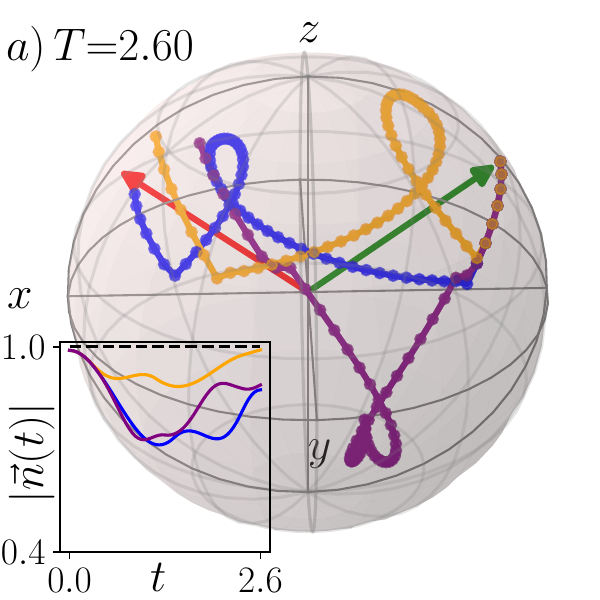}
\includegraphics[width=.235\textwidth]{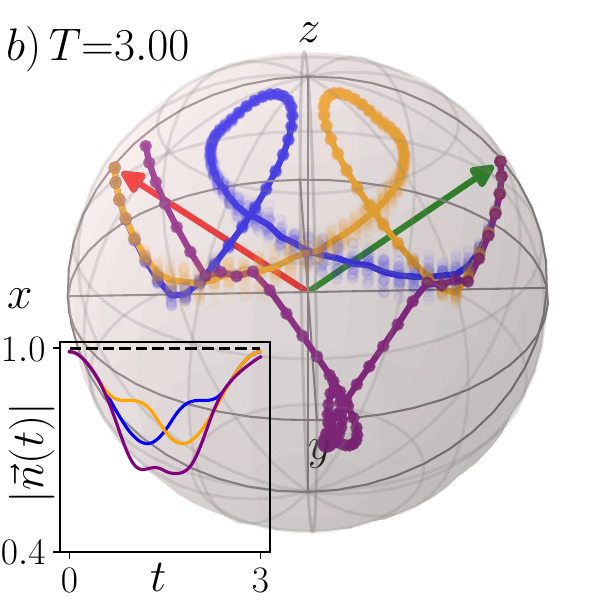}
\includegraphics[width=.235\textwidth]{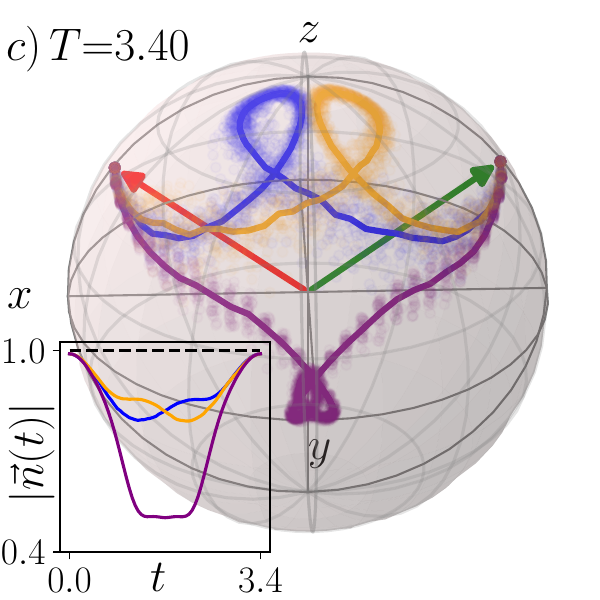}
\includegraphics[width=.235\textwidth]{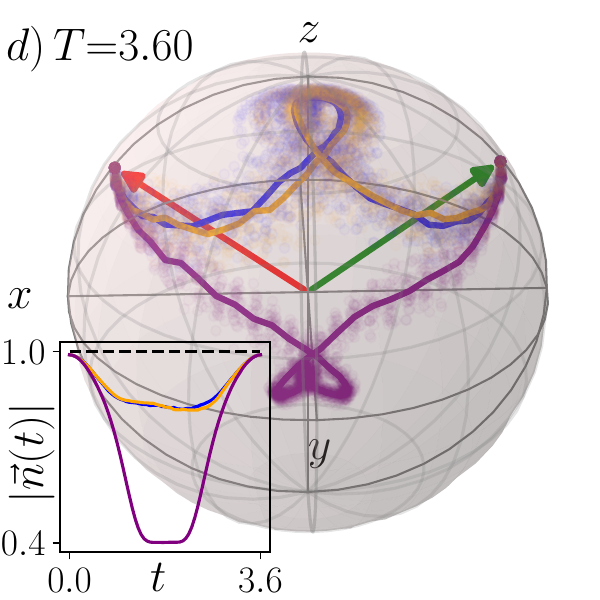}
\caption{
Topological transitions are experimentally accessible in modern-day experiments. The figure shows Bloch sphere trajectories in the reduced density matrix obtained from protocols sampled by the Langevin-Monte Carlo (LMC) algorithm. The green (red) arrow indicates the initial (target) state.
Solid lines represent \emph{averages} over three LMC runs, each sampling one of the three connected components $C_+,C_-,C_3$ of the optimal level set (colored in blue, orange and purple, respectively);
low-opacity dotted trajectories represent \emph{single} protocols and serve to visualize protocol fluctuations.
\textbf{a)} $T{=}2.6{>}T_\text{sb}$: blue and orange trajectories correspond to the isolated optimal protocols $s_+,s_-$; the purple curve $s_3$ is a local trap.
\textbf{b)} $T{=}3.0{>}T_\text{QSL}$: optimal level set consists of two connected components $C_+,C_-$, each containing infinitely many optimal protocols (cf.~blue and orange clouds surrounding the solid lines).
\textbf{c)} $T{=}3.4{>}T_\text{t+}$: the orange and blue clouds are still well-separated; the purple protocol, now also optimal, gives rise to the third connected component $C_3$.
\textbf{d)} $T{=}3.6{>}T_\text{t--}$: orange and blue trajectories overlap and the corresponding clouds are indistinguishable: the previously distinct connected components $C_+,C_-$ have merged and formed $C_\pm$.
Insets: the purple trajectories reach the largest entanglement entropy as the associated norm $\abs{n(t)}$ drops to ${\approx}0.5$ during the evolution.
}
\label{fig:bloch_sphere}
\end{figure}

Figure~\ref{fig:bloch_sphere} shows three optimal trajectories of $\vec n_1(t)$ generated by three distinct protocols, for $T{=}2.6,3.0,3.4,3.6$. Each solid line shows the \emph{average} trajectory over a single LMC run, sampled from a single connected component among $C_+, C_-, C_3$ (blue, orange, and purple, respectively).
Notice that, for $T{<}T_\text{t+}$, $C_3$ consists of an isolated local trap (i.e., its trajectory does not reach the target state) and hence it does not belong to the optimal level set. 
To visualize fluctuations within each connected component, we plot a subset of protocols in each LMC run using half-transparent dots; in this way, fluctuations within the same connected component generated by the LMC dynamics are visible as a ``cloud'' of trajectories distributed around the solid line of the same color.

For $T{=}2.60{<}T_\text{QSL}$, the dotted trajectories coincide with the solid line, in agreement with optimal protocols being isolated.
For $T{=}3.00{\approx}T_\text{QSL}$, trajectories from the $C_+,C_-$ connected components (blue and orange curves) connect the initial state (green arrow) with the target state (red arrow), in accord with the vanishing infidelity associated with $C_+,C_-$ past $T_\text{QSL}$. Here, dotted trajectories start deviating from the solid line, and a semi-transparent cloud of trajectories in the Bloch sphere is visible around it.
For $T{>}T_\text{t+}$, component $C_3$ has vanishing infidelity, and the purple trajectories now connect the initial and final states. 
Nevertheless, the three solid lines and the surrounding clouds of trajectories are distinguishable from each other: LMC cannot jump between the different connected components for large enough $\beta$. 
Finally, for $T{>}T_\text{t--}$ the blue and orange curves are indistinguishable from one another: $C_+,C_-$ have merged into the single connected component $C_\pm$.
For $T{\ge}T_\text{t--}$, we assign to LMC trajectories in the $C_\pm$ connected components orange or blue colors depending on the average magnetization of protocols in each run.

Different connected components may possess substantially different physical features. For example, in the insets of Fig.~\ref{fig:bloch_sphere} we show the norm $\abs{n(t)}$ of the Bloch sphere vectors. 
The entanglement entropy $S_E=-q\log q - (1-q)\log(1-q)$ with $q=(\abs{n}+1)/2$ is different in the three cases. In particular, the minimum value $\abs{n}{\approx}0.5$ reached by $C_3$ trajectories (purple), is half the value reached by $C_+,C_-$ trajectories (orange and blue). We expect that, if present, noise in the form of decoherence or depolarization may reveal some of these trajectories as more robust than others.

As supplemental material \cite{Supplemental}, we provide an animation showing three optimal trajectories in the Bloch sphere, which represents the evolution of the one-qubit reduced density matrix. 
The blue and orange trajectories are generated by optimal protocols with total duration $T{=}3.4$. They belong to the different connected components $C_+,C_-$ of the optimal level set (they cannot be continuously deformed into each other while preserving vanishing infidelity). 
The black trajectory has a total duration $T{=}3.6$ and its dynamics is governed by the LMC algorithm. 
The movie demonstrates that for $T{=}3.6$ the blue and orange connected components have merged into a single connected component: $C_+ C_- {\to} C_\pm$. This is proven by the black trajectory that, through approximate continuous deformations (LMC dynamics with $\beta{=}10^6$), moves between the ``orange" and ``blue" regions of $C_\pm$.

\pagebreak
\clearpage

\setcounter{page}{1}

\begin{widetext}
\begin{center}
\textbf{Supplemental Information for \\
``Topological phase transitions in a constrained two-qubit quantum control landscape"}
\end{center}
\end{widetext}

\section{Local trap in the quantum control landscape}
\label{app:trap}

For $2.5{\lesssim}T{\lesssim}3.4$ we observe a number of LMC runs stabilizing around a sub-optimal protocol (local trap), which we called $s_3$. In this section, we demonstrate that $s_3$ gives rise to the third connected component $C_3$ in the optimal level set, for $T{\ge}T_\text{t+}$.

\begin{figure}
\centering
\includegraphics[width=.45\textwidth]{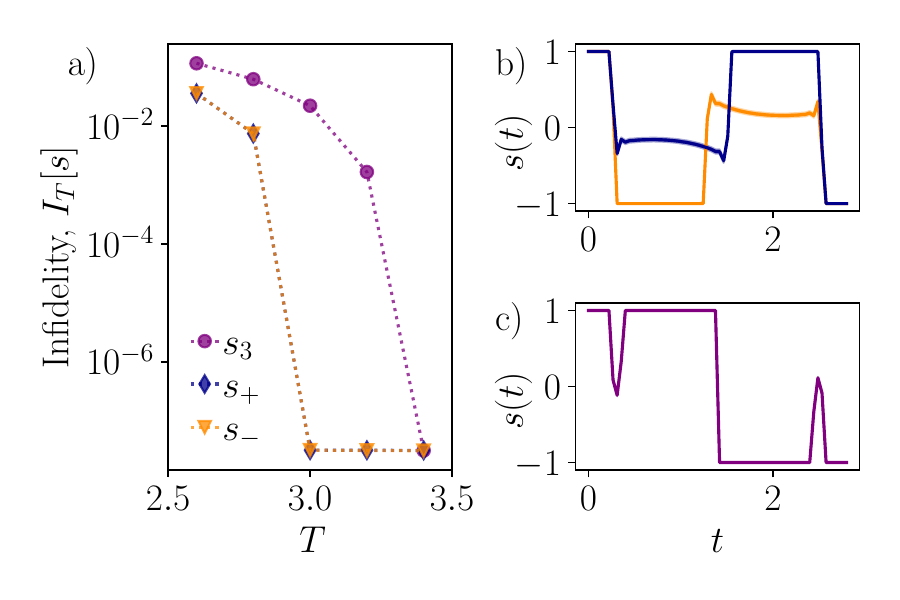}
\caption{
    A local trap $s_3$ appears in the landscape for $T{\in}[2.6,3.2]$.
    \textbf{a)} Infidelity associated with the locally stable protocols $s_+,s_-,s_3$ found via LMC (in blue, orange and purple, respectively) shows that $s_3$ is a local trap. 
    \textbf{b,c)} Visualization of protocols $s_+,s_-,s_3$ at $T{=}2.80$. Contrary to $s_+$ and $s_-$, $s_3$ does not break the control problem symmetry $s(t){\leftrightarrow} {-}s(T{-}t)$.
}
\label{fig:trapF}
\end{figure}

On the left side of Fig.~\ref{fig:trapF} we show the infidelity curve associated with the three locally-stable protocols found by LMC, $s_+,s_-,s_3$ (colored in blue, orange, and purple, respectively). The purple curve, lying above the blue/orange curves, reveals the protocol $s_3$ as a local trap in the quantum control landscape. 
From this curve, we also see that $s_3$ becomes globally optimal around $T{\approx}3.4$; 
we deduce that $C_3$, the third connected component appearing for $T{\ge}3.4$, originates from the local-trap $s_3$.

On the right side of Fig.~\ref{fig:trapF} we display an example of the three locally stable protocols found by LMC for $T{=}2.8$. Among the three protocols, $s_\pm$ are globally optimal whereas $s_3$ is sub-optimal; in particular, $I[s_\pm]{\approx}0.01$ to be compared with $I[s_3]{\approx}0.06$.

The fraction of LMC runs trapped around $s_3$ is reduced by introducing a Simulated Annealing (SA) schedule, during the thermal relaxation stage of the LMC dynamics \cite{beato2025toward}. Using SA we observe that the probability of finding LMC stuck in the local trap is less than 1/10, for $L{=}64$ (cf.~Sec.~\ref{sec:perforamance} and \ref{sec:basins}.
This probability decreases for a larger number of protocol discretization steps $L$, because the distribution of the initial protocol configurations concentrates around the zero protocol $s(t_i){=}0$, for $L{\to}\infty$.
This is due to the random initialization of the initial protocol in LMC, where each site $s(t_i)$ is drawn from a uniform distribution in $[0,1]$. Since each site corresponds to a time step $\delta t {=} T/L$, for $\delta t {\ll} \max_t \norm{\hat H(t)}^{-1}$ the distribution of initial protocols concentrates around the zero protocol.

\begin{figure}
    \centering
    \includegraphics[width=1.\linewidth]{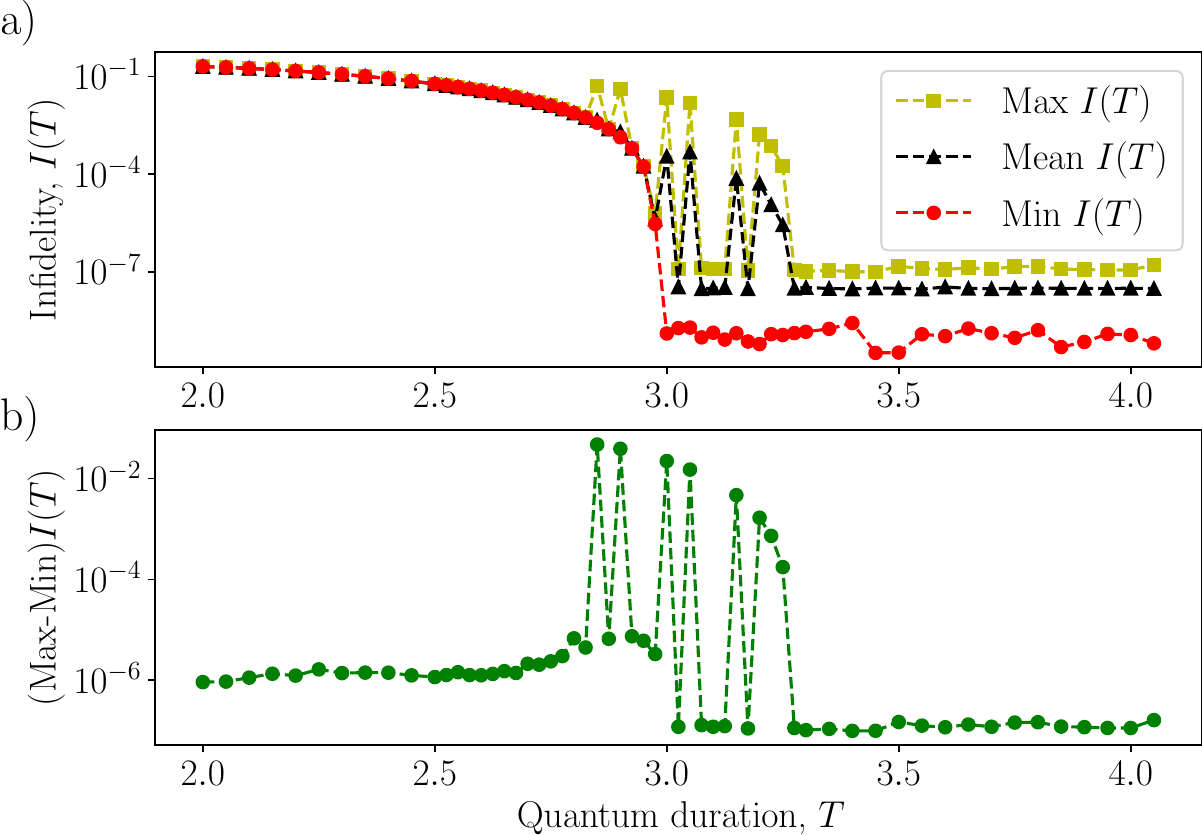}
    \caption{
    $\textbf{a)}$ Maximum, minimum and mean infidelity obtained at the end of the optimization performed by the Langevin-Monte Carlo optimization with simulated annealing, for $T\in[2.0,4.0]$.
    $\textbf{b)}$ Difference between the maximum and minimum infidelity obtained.
    For this analysis, we consider a set of $2^{10}$ independent LMC runs at each $T$.
    We see that the local trap $s_3$ impacts the optimization performance around the quantum speed limit, $T_\text{QSL}\approx3.0$.
    }
    \label{fig:mmm}
\end{figure}

\section{Performance of optima found in practice}
\label{sec:perforamance}

In this section, we comment on the performance of minima found by the Langevin-Monte Carlo algorithm used in combination with simulated annealing (LMC+SA). 

In the upper plot of Fig.~\ref{fig:mmm}, we show the maximum, minimum, and mean infidelity reached by $1024$ LMC runs, with $L=64,\beta=10^6$, for $T\in[2.0,4.0]$.
In the lower plot, we show the difference between the maximum and the minimum infidelity.
From this result, we notice that a significant fraction of LMC runs are trapped in a sub-optimal minimum only around the quantum speed limit $T_\text{QSL}$. The precise value of the fraction depends on the schedule of Simulated Annealing (SA).
In the worst case, the local trap $s_3$ causes a deviation from the best known infidelity of order $\sim10^{-1}$.

\begin{table}
\centering
\begin{tabular}{cccccc}
\toprule
$T$ & $N_{C_+}$ & $N_{C_-}$ & $N_{C_\pm}$ & $N_{C_3}$ & $R$\\
\midrule
3.40    & 28     & 32     &  & 4 & 64 \\
3.75    &   &   & 58     & 6 & 64 \\
4.00    &   &   & 62     & 2 & 64 \\
\bottomrule
\end{tabular}
\caption{
    Number of LMC runs in each connected component, for different quantum duration $T$ (out of a total of $R=64$ LMC runs).
    We observe that the connected component $C_3$ traps approximately $1/10$ of the LMC runs; its basin of attraction is significantly smaller than the ones of $C_+,C_-,C_\pm$.
}
\label{tab:basins}
\end{table}

\section{Basins of attraction analysis}
\label{sec:basins}

In this section, we extract the number of LMC runs in each connected component from the distance distribution $P(d_\text{avg})$.

We focus on the intervals $T\in[3.3,3.5]$ and $[3.5,4.0]$, where the system has three ($C_+,C_-,C_3$) and two ($C_\pm,C_3$) connected components, respectively.
The results for $T=3.40, 3.75, 4.00$ are summarized in Table \ref{tab:basins}.
We see that the connected component $C_3$ traps about $\sim1/10$ of all LMC runs, while the rest is attracted to $C_\pm$ or $C_+ \cup C_-$.
Consistent with their symmetry, we observe $N_{C_+} \approx N_{C_-}$.

In the following, we discuss the calculations leading to the results presented in Table \ref{tab:basins}.
$P(d_\text{avg})$ is the distribution of the average distance $d_\text{avg}$ between protocols in two different LMC runs.
Hence, we use the cumulative probability associated with single peaks to quantify the number of LMC runs in each connected component. 
In particular, if $P(C)$ is the cumulative distribution associated with the self-distance peak of connected component $C$, we extract the number of LMC runs in that connected component, $N_C$, from the condition
\begin{equation*}
    Z P(C) = N_C(N_C-1)/2,
\end{equation*}
where $Z=R(R-1)/2=2016$ is the number of the pairs among the $R=64$ different LMC runs we consider.
As the self-peaks of $C_+$ and $C_-$ overlap, this argument only gives us the number of runs in \emph{either} $C_+$ \textit{or} $C_-$, namely $N_{C_+ \cup C_-}$. 
Therefore, to obtain $N_{C_+}$ and $N_{C_-}$, we combine information from the cumulative probability of the self and cross distance peaks, $P(C_+)+P(C_-)$ and $P(C_+,C_-)$ respectively. Namely,
\begin{align*}
    Z (P(C_+)+P(C_-)) &= \frac{N_{C_+}(N_{C_+}{-}1)}{2} + \frac{N_{C_-}(N_{C_-}{-}1)}{2}. \\
    Z P(C_+,C_-) &= N_{C_+}N_{C_-}.
    \end{align*}
This system of two algebraic equations is symmetric under the exchange of $N_{C_+}$ and $N_{C_-}$, so that this analysis does not distinguish between $C_-$ and $C_+$. 
As discussed in the main text, the two connected components are distinguished by the ``magnetization'' order parameter.

\section{Distribution of protocol distances \texorpdfstring{$P(d_\text{avg})$}{P(davg)} }
\label{app:d_a_scalings}

This section shows how the distribution $P(d_\text{avg})$ defined in Eq.~\eqref{eq:d_a} is affected by the LMC parameters.

\begin{figure}[t!]
	\centering
	\includegraphics[width=.45\textwidth]{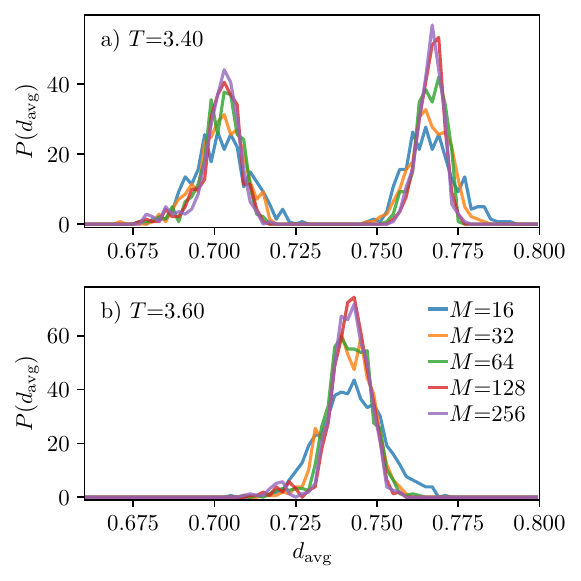}
	\caption{
		Density distribution $P(d_\text{avg})$ of the \emph{average} distance before (panel a) and after (panel b) the $T_\text{t--}$ topological CLPT.
        The behavior of $P(d_\text{avg})$ is shown as a function of the number of protocols $M$ considered for evaluating $d_\text{avg}$.
        We observe that the width of the peaks shrinks with $M$. The peaks are well separated already at $M{=}16$ for the particular values of $T$ considered.
	}
	\label{fig:d_a_dist_scaling}
\end{figure}

LMC collects $M{=}2^{12}$ protocols, waiting for $\Delta n{=}2^{14}$ LMC iterations between two samples to guarantee uncorrelated sampling (each iteration corresponds to $L$ attempted local updates).
To produce Fig.~\ref{fig:fig2}a, we extract $2^8$ protocols from each LMC run. In Fig.~\ref{fig:d_a_dist_scaling}, we show that the average distance ($d_\text{avg}$) distribution does not require a larger number of sampled protocols to separate different peaks, for the particular discretization of $T{\in}[0,4]$ chosen in Fig.~\ref{fig:fig2}a.

\begin{figure*}
\centering
\includegraphics[width=.3\textwidth]{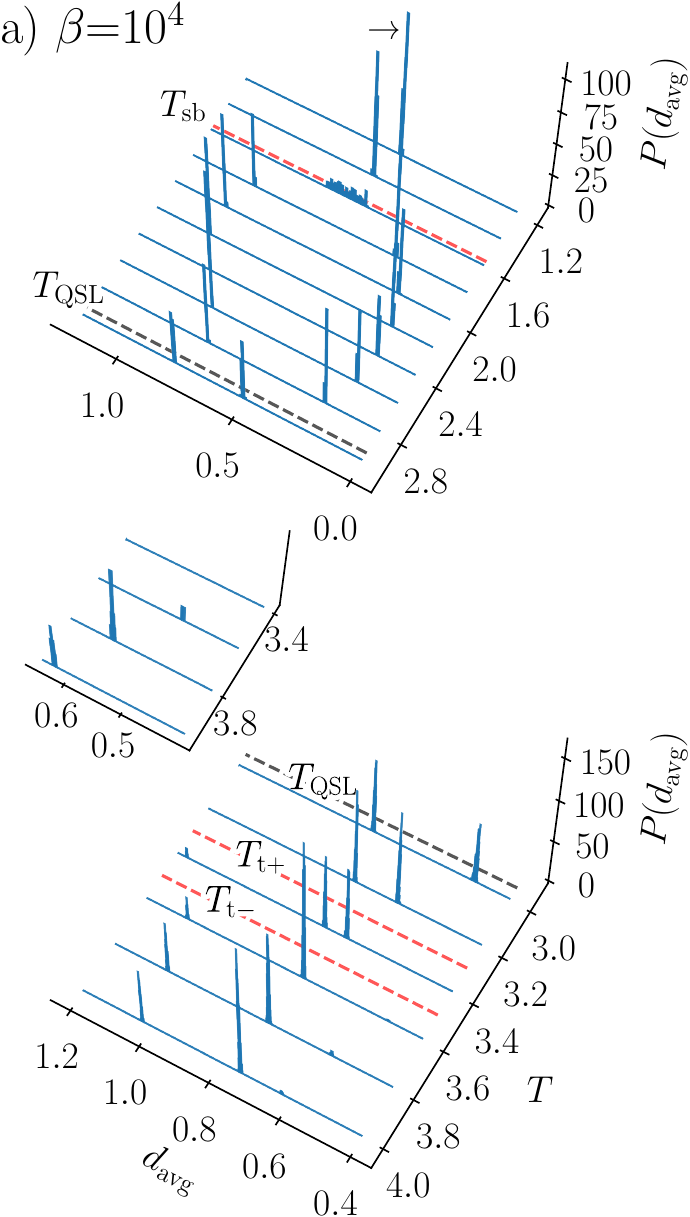}
\includegraphics[width=.3\textwidth]{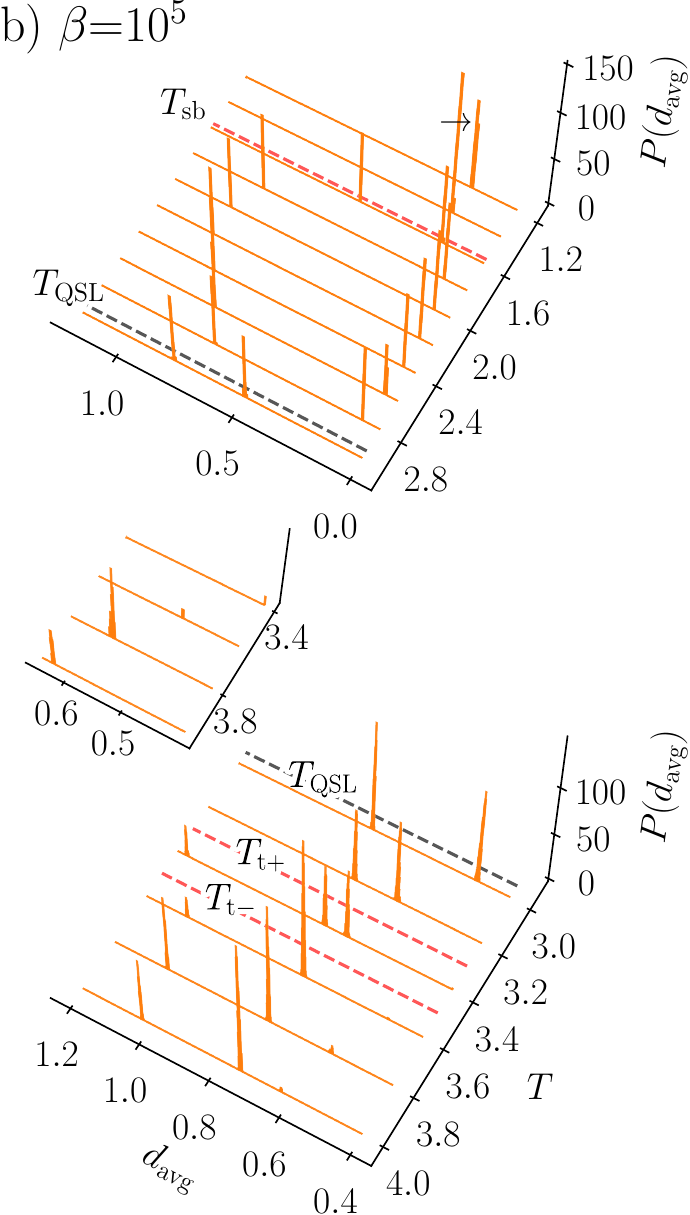}
\includegraphics[width=.3\textwidth]{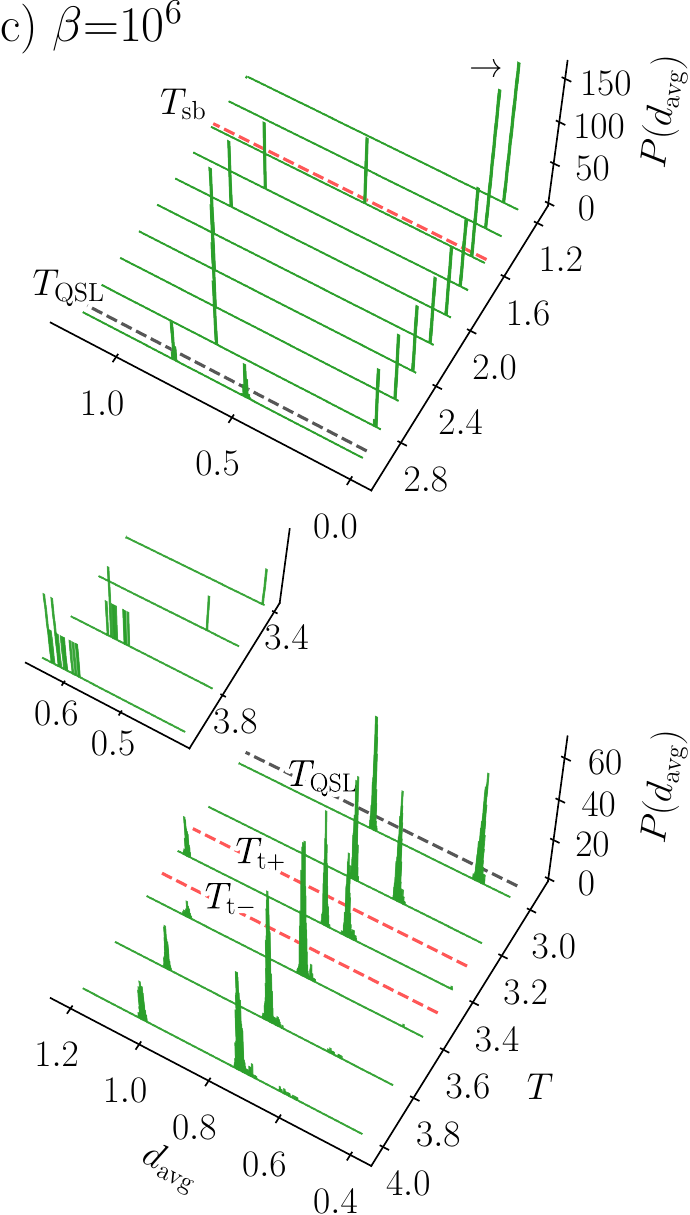}
\caption{
    Density distribution $P(d_\text{avg})$ of \emph{average} distance as a function of the LMC inverse-temperature $\beta{=}10^4,10^5,10^6$ ($L{=}64$).
	The only quantitative change in $P(d_\text{avg})$ in the three panels occurs for $T{\in}[0,3.0]$. The peak located in $d_\text{avg}{\approx}0.25$ (indicated by the arrow) shifts towards $d_\text{avg}{=}0$, as $\beta$ increases.
    The main features of $P(d_\text{avg})$ are already visible at $\beta{=}10^4$ and are stable in the range $\beta{\in}[10^4,10^6]$. We deduce that results obtained with $\beta{=}10^6$ represent a good approximation of the situation in limit $\beta{\to}\infty$.
}
\label{fig:d_a_dist_Btpscaling}
\end{figure*}

In Fig.~\ref{fig:d_a_dist_Btpscaling} we show the scaling in $\beta$ of the distribution of distance $P(d_\text{avg})$.
The only appreciable change in $P(d_\text{avg})$ involves the peak closest to $d_\text{avg}{=}0$, for $T{\le}3.0$. This peak originates from pairs of LMC runs sampling in the same connected component ($C_0$ for $T<T_\text{sb}$; $C_+,C_-$ for $T_\text{sb}{<}T{<}T_\text{QSL}$) consisting of isolated optimal protocols. Decreasing $\beta$ reduces thermal fluctuations around the isolated optimal protocols, thus shifting this peak towards $d_\text{avg}{=}0$.

\begin{figure*}
\centering
\includegraphics[width=.9\textwidth]{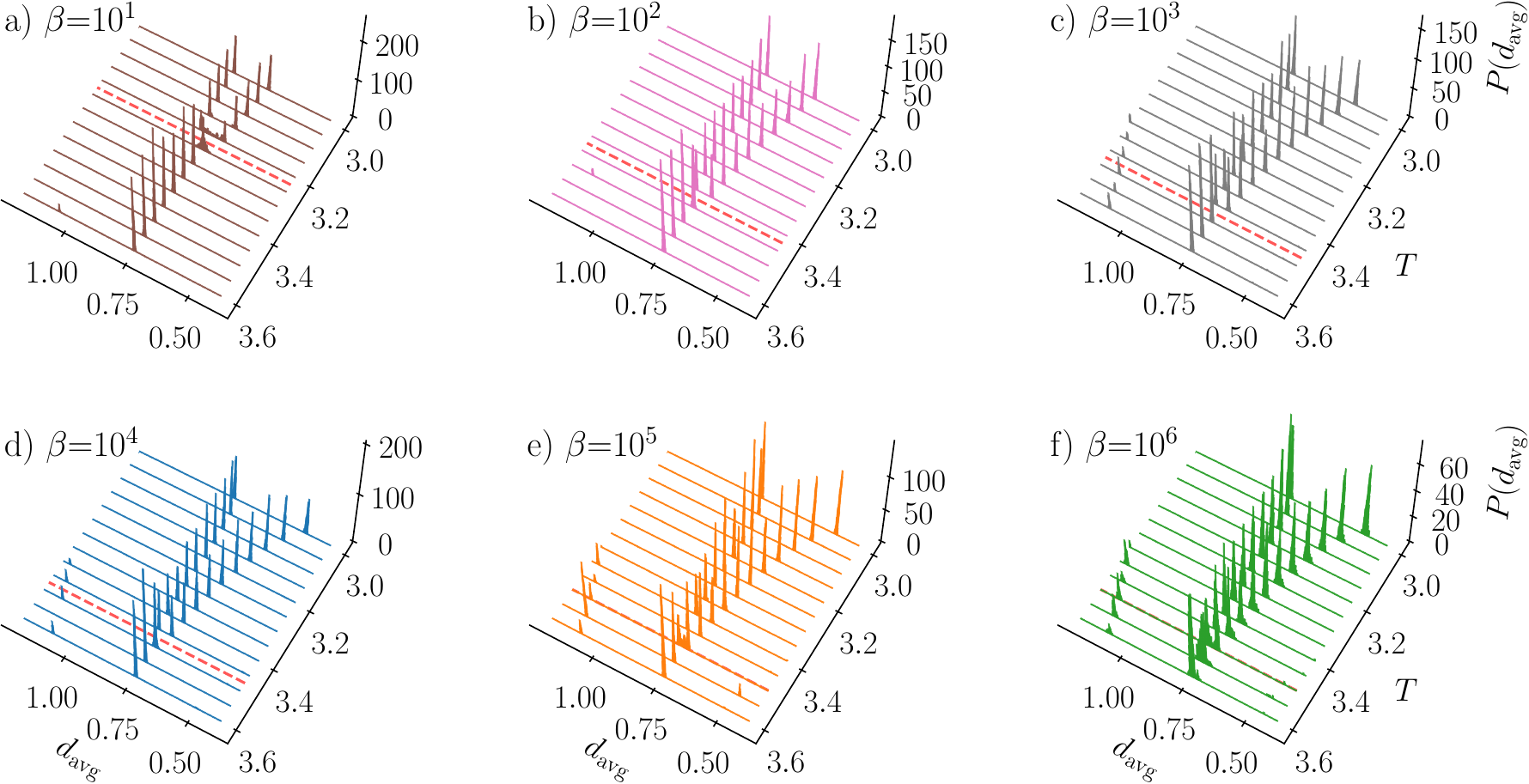}
\caption{
    Density distribution $P(d_\text{avg})$ of \emph{average} distance around the $T_\text{t--}{\approx}3.5$ topological CLPT, as a function of the LMC inverse-temperature $\beta{=}10^1,10^2,\dots,10^6$ ($L{=}64$).
    Effects of thermal fluctuations on the $T_\text{t--}$ transition are relevant for $\beta{\lesssim}10^2$ (cf.~red dashed line).
    From this result, we estimate the infidelity barrier separating the two connected components $C_+,C_-$ (see main text).
}
\label{fig:d_a_dist_lowBtp}
\end{figure*}

Interestingly, for $T{>}T_\text{QSL}$ and $L{=}64$ thermal fluctuations up to $\beta{\approx}10^4$ do not change the location of the $T_\text{t--}$ transition. Nevertheless, thermal fluctuations will eventually allow hopping between $C_+,C_-$ for small enough $\beta$.
To test this idea, and obtain a numerical estimate of the infidelity barrier separating the two connected components, in Fig.~\ref{fig:d_a_dist_lowBtp} we plot $P(d_\text{avg})$ for $3.0{\le}T{\le}3.6$ and $\beta{=}10^1,10^2,\dots,10^6$. 
We observe that the location of the topological transition is severely affected only for $\beta{\lesssim}10^2$. 
By comparing with the acceptance probability $\exp(-\beta L \Delta I)$ in the LMC algorithm, we deduce that the two connected components are separated by a barrier of order $\Delta I {\approx} 10^{-3}$ for $T{=}3.4$.

\begin{figure*}
\centering
\includegraphics[width=.3\textwidth]{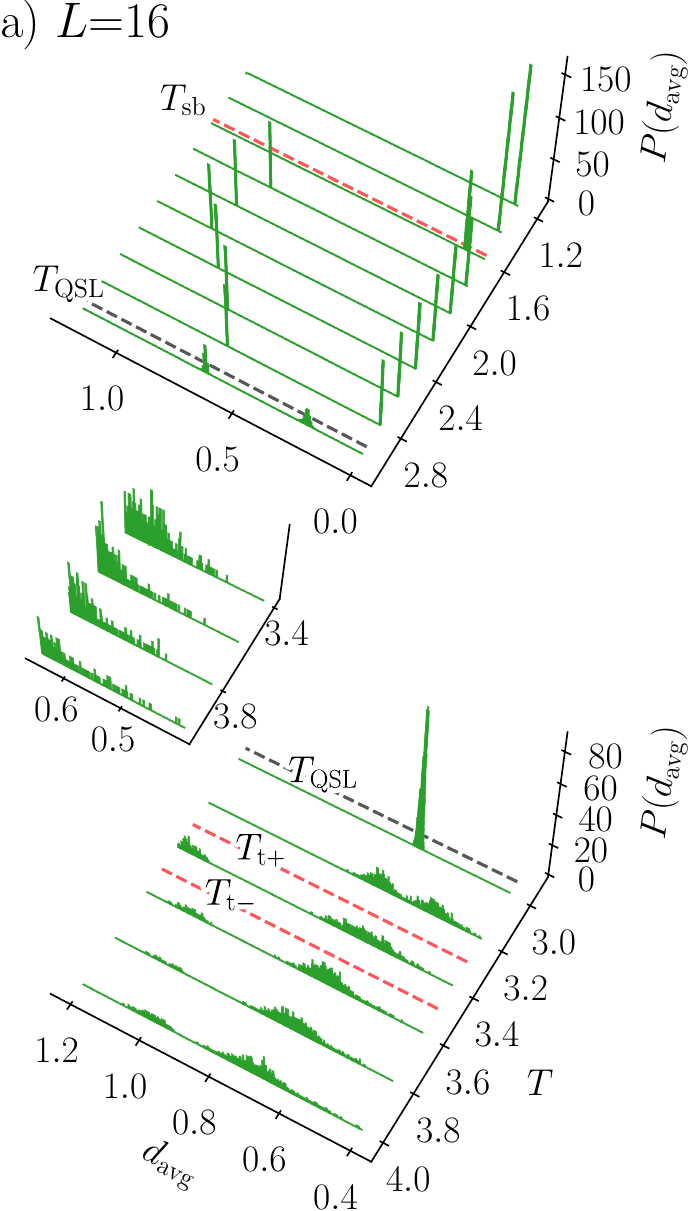}
\includegraphics[width=.3\textwidth]{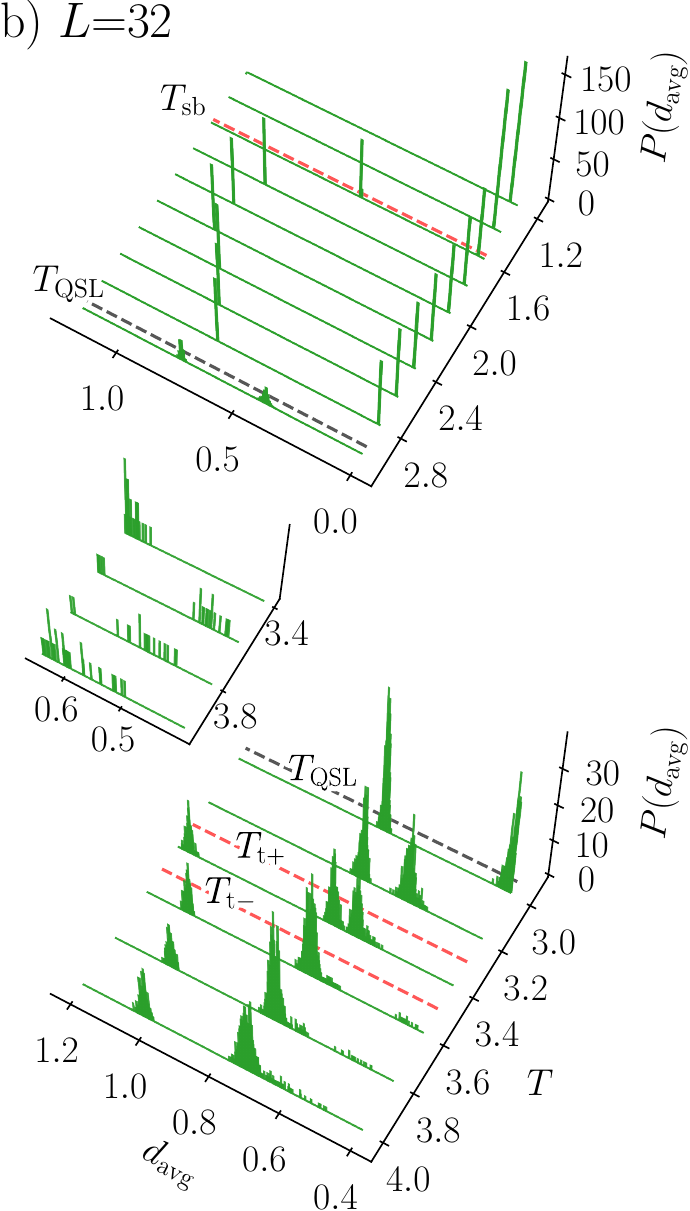}
\includegraphics[width=.3\textwidth]{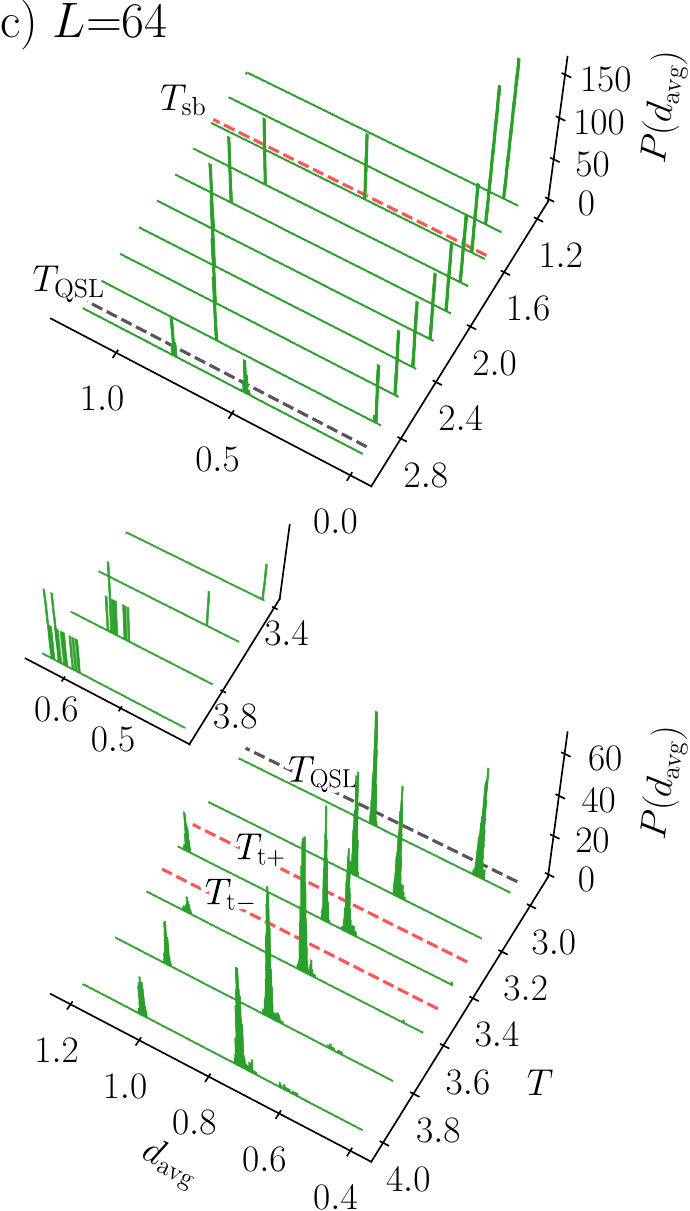}
\caption{
    Density distribution $P(d_\text{avg})$ of \emph{average} distance shown as a function of the number of protocol steps $L{=}16,32,64$ ($\beta{=}10^6$).
    $L$ plays an important role in the correct identification of the topological CLTP in $T_\text{t--}$, as the width of the peaks in $P(d_\text{avg})$ past the quantum speed limit depends on $L$, for a fixed number of sampled protocols. For example, for $L{=}16$ at $T{=}3.4$ there is a single peak in $d_\text{avg}{\approx}0.7$, while for $L{=}64$ there are two well-separated peaks. 
}
\label{fig:d_a_dist_Lscaling}
\end{figure*}

Finally, in Fig.~\ref{fig:d_a_dist_Lscaling} we show the scaling in $L$ of the distribution $P(d_\text{avg})$. 
For $T{<}3.0$ there are no appreciable changes. For $T{\ge}3.0$ the peaks in the distribution broaden as $L$ is \emph{decreased}. This phenomenon has important consequences for the location of the $T_\text{t--}$ transition: for $L{=16}$ the two peaks are indistinguishable from one another already at $T{=}3.4$; in turn, between $L{=}32,64$ there are no noticeable finite-size effects.

\section{Alternative distance distributions: \texorpdfstring{$P(d_\text{set})$}{P(dset)} and \texorpdfstring{$P(d_\text{prt})$}{P(dprt)} }
\label{app:d_p_d_min}

In the main text, the average distance $d_\text{avg}$ between pairs of protocols is considered. In this section, we present two alternative distance definitions, show that they lead to the same conclusions and compare the associated computational costs.

The \emph{set} distance between $\mathcal S_\alpha$ and $\mathcal S_\beta$ is defined as 
\begin{equation}
    d_\text{set}(\mathcal S_\alpha, \mathcal S_\beta) = \min_{i,j} d(s^\alpha_i, s^\beta_j).
    \label{d_set}
\end{equation}
whereas the \emph{average protocol} distance as
\begin{equation}
    d_\text{prt}(\mathcal S_\alpha, \mathcal S_\beta) = 
    d\qty(
    \frac1M \sum_{i=1}^M s^\alpha_i, 
    \frac1M \sum_{j=1}^M s^\beta_j
    ).
    \label{eq:d_p}
\end{equation}

\begin{figure*}
\centering
\includegraphics[width=.3\textwidth]{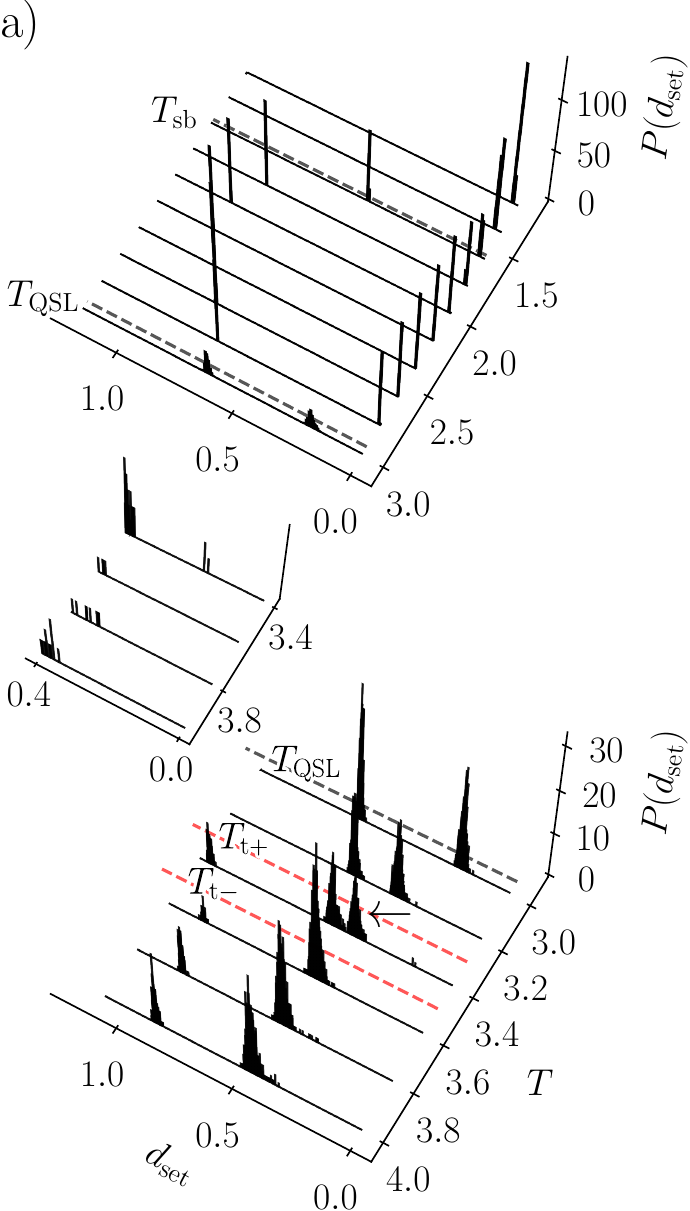}
\includegraphics[width=.3\textwidth]{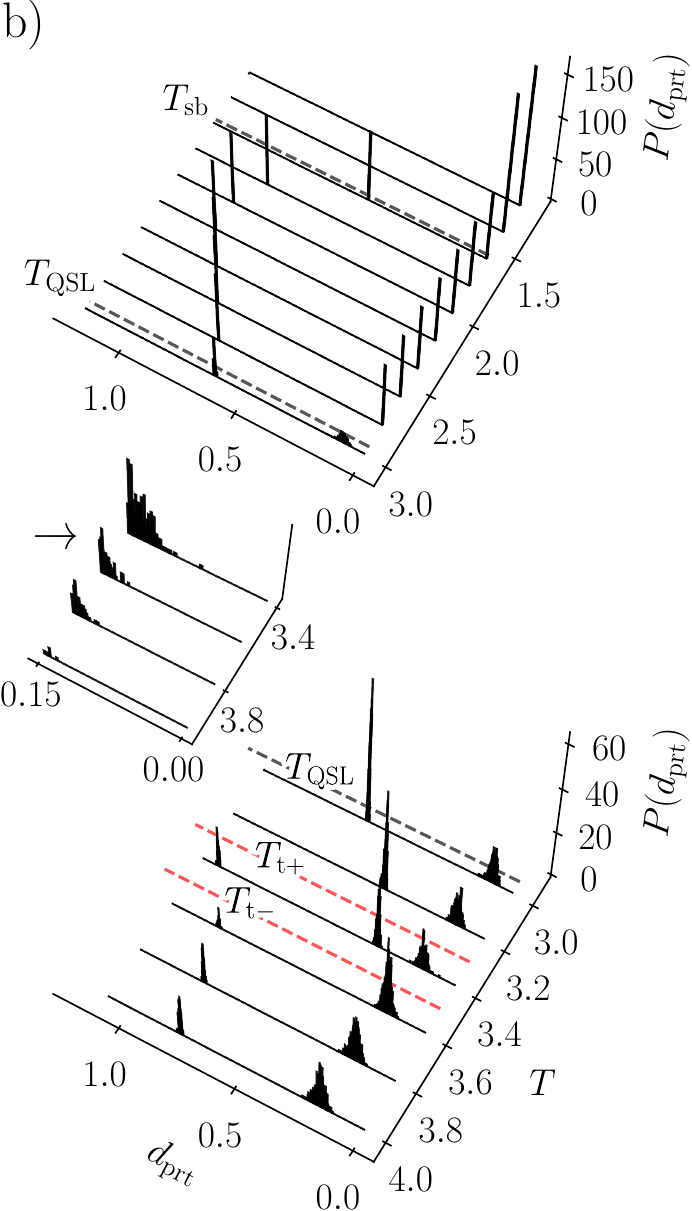}
\caption{
    Density distributions $P(d_\text{set}),P(d_\text{prt})$ of the \emph{set} and \emph{average protocol} distances, respectively ($L{=}64,\, \beta{=}10^6$).
    \textbf{a)} In $P(d_\text{set})$, the location of the $T_\text{t--}$ topological CLPT can be identified with less precision than $P(d_\text{avg})$: peaks in $P(d_\text{set})$ have a \emph{larger} width than $P(d_\text{avg})$ and they partially overlap already for $T{=}3.4$ (indicated by the arrow).
    \textbf{b)} $P(d_\text{prt})$ displays the same information content as $P(d_\text{avg})$ with the exception of the two overlapping peaks around $d_\text{prt}{\approx}0.2$ for $T{\in}[3.4,4.0]$ (indicated by the arrow).
}
\label{fig:ds_dist}
\end{figure*}

In Fig.~\ref{fig:ds_dist} we show the distributions $P(d_\text{prt}),P(d_\text{set})$ as a function of $T{\in}[0,4]$. $P(d_\text{avg}),P(d_\text{set}),P(d_\text{prt})$ exhibit the same qualitative features, and we conclude that the results presented in the main text are independent of the particular distance definition.

Nevertheless, let us remark on two important differences between $P(d_\text{prt}),P(d_\text{set})$ and $P(d_\text{avg})$ (for the same set of LMC parameters). 
First, $P(d_\text{prt})$ does not always allow us to distinguish all the peaks associated with the distance network in Fig.~\ref{fig:fig2}b. Consequently, $P(d_\text{prt})$ may lead to a wrong reconstruction of connected components in the optimal level set.
For example, in contrast to Fig.~\ref{fig:fig2}a, $P(d_\text{prt})$ exhibits only \emph{three} peaks for $T{\in}[3.4,4.0]$; in particular, the two peaks $d_\text{prt}(C_3,C_3)$ and $d_\text{prt}(C_\pm,C_\pm)$ around $d_\text{prt}{\approx}0.2$ (``self-distances'') are not distinguishable from each other.
Second, $d_\text{set}$ is more sensitive to fluctuations within the optimal level set than $d_\text{avg}$.
For this reason, $P(d_\text{set})$ requires a larger number $M$ of protocols sampled from each LMC run to reach the same resolution.
In particular, we see in Fig.~\ref{fig:ds_dist} that the peaks of $P(d_\text{set})$ are significantly broader than $P(d_\text{avg})$ and $P(d_\text{prt})$, for the particular number of LMC protocols considered ($M{=}2^{12}$).
For example, at $T{=}3.4$ the peaks around $d_\text{set}{\approx}0.75$ are partially overlapping.

As a last comment, we remark that the three distance definitions require different computational costs: while $d_\text{set},d_\text{avg}$ require in principle to consider all pairs of protocols in the sets $\{\mathcal S_\alpha\}$ (an $\order{M^2}$ operation) $d_\text{prt}$ only needs the evaluation of the distance between two protocols.
In practice, in $P(d_\text{set})$ we considered all $M{=}2^{12}$ protocols sampled in each LMC run while in $P(d_\text{avg})$ it suffices to take into account only a subset of $2^8$ protocols among the $M$ sampled.

The results shown in this section suggest that $d_\text{avg}$ is the most convenient distance to investigate the topology of the optimal level set in the two-qubit control problem.

\section{Topological transitions without control symmetry breaking}
\label{sec:asymmetry}

In this section, we demonstrate the independence of topological CLPTs from the control symmetry $s(t) \leftrightarrow -s(T-t)$, possessed by the control problem we discuss in the main text.
To this end, let us change the control problem and explicitly break its symmetry.
One simple way to do so is to choose the initial state $\ket{\psi_0}$ as the ground state of the Hamiltonian 
\begin{align}
    \hat H_0
    &= -J \hat S_1^z \hat S_2^z - h_x (\hat S_1^x + \hat S_2^x)  \notag\\
    &- h_y (\hat S_1^y + \hat S_2^y) - h_z (\hat S_1^z + \hat S_2^z) 
    \label{eq:H_1}
\end{align}
with $J{=}2,\, h_x{=}2,\, h_y{=}0.5,\, h_z{=}1$. 
For the final state, $\ket{\psi_*}$, and the quantum evolution, $\hat H(t)$, we use the same as in the main text (cf.~Eq.~\eqref{eq:H}).
Compared to the control problem discussed in the main text, the term $h_y \hat S_y$ in $H_0$ explicitly breaks the symmetry $s(t) \leftrightarrow -s(T-t)$ of the infidelity $I_T[s]$ \cite{Bukov18_Broken}.

\begin{figure*}
    \centering
    \includegraphics[width=.3\linewidth]{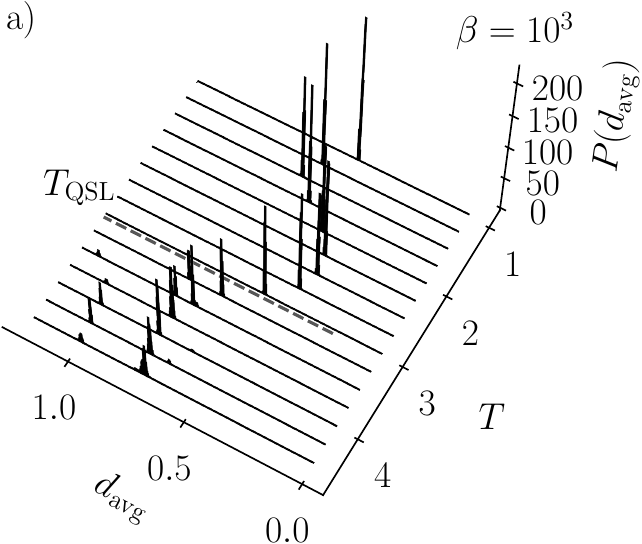}
    \includegraphics[width=.3\linewidth]{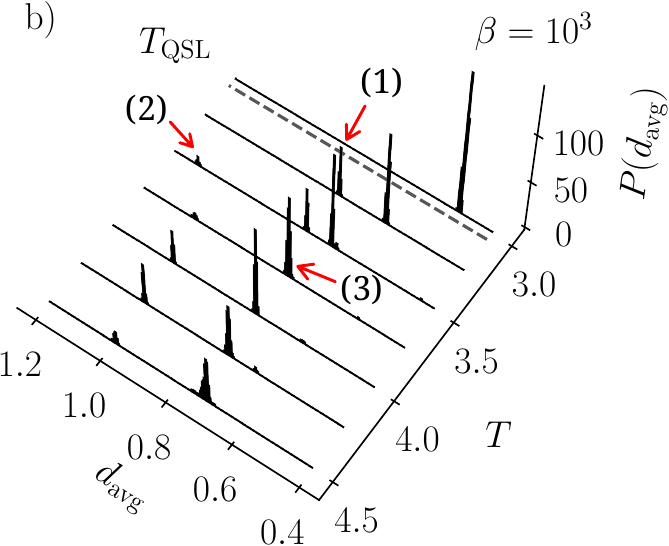}
    \includegraphics[width=.3\linewidth]{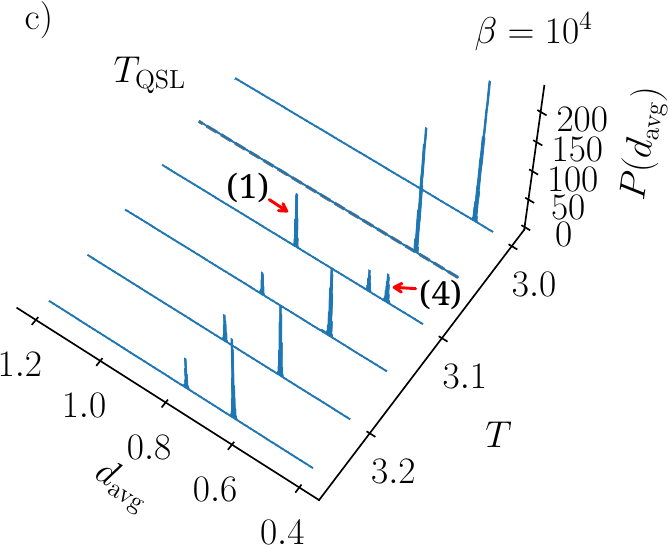}
    \caption{
    Example of a quantum control problem whose infidelity landscape violates the control symmetry $s(t) \leftrightarrow -s(T-t)$ and which nonetheless exhibits topological CLTPs.
    Here, the initial state $\ket{\psi_0}$ is the ground state of the Hamiltonian \eqref{eq:H_1}.
    \textbf{a)} 
    The distance distribution $P(d_\text{avg})$ exhibits a single peak up to $T = 3.25 > T_\text{QSL} \approx 3.05$. Hence, no symmetry-breaking transition occurs in this model.
    \textbf{b)}
    Plot of $P(d_\text{avg})$ in the region $T\in[3.25,4.5]$, where topological transitions occur.
    We observe (i) new connected components emerging at $T \approx 3.25,3.50$ (arrow \textbf{1} and \textbf{2}) and (ii) two connected components merging at $T\approx3.75$ (arrow \textbf{3}).
    \textbf{c)} 
    Plot of $P(d_\text{avg})$ in the region $T\in[3.00,3.25]$, where the second connected component emerges and the asymmetry is visible.
    The asymmetry between two connected components affects their size and is visible in $P(d_\text{avg})$: for $T=3.10$, the peaks corresponding to the self-distance of the two components do not overlap (arrow \textbf{4}). 
    For $T>3.10$, the two self-distance peaks in $P(d_\text{avg})$ cannot be resolved at the level of the resolution provided by our numerical simulations.
    LMC parameters: $L=64,\, \beta=10^3,\, \Delta n=2^8$ in \textbf{a)} and \textbf{b)}; $\beta=10^4,\, \Delta n=2^{14}$ in \textbf{c)}.
    }
    \label{fig:asymmetry}
\end{figure*}

In Fig.~\ref{fig:asymmetry}, we report the corresponding distance distribution $P(d_\text{avg})$ for $T\in[0,4.5]$. As expected, we observe the same qualitative features as in the example discussed in the main text, except for the symmetry-breaking transition: a single connected component is present in the landscape up to $T = 3.10 > T_\text{QSL} \approx 3.05$.
In particular, we still observe the emergence of new connected components at $T\approx3.10$ and $T\approx3.50$ and the merging of two connected components at $T\approx3.75$ (cf.~Fig.~\ref{fig:asymmetry}b).
The asymmetry between two connected components affects their size and is thus visible in $P(d_\text{avg})$: for $T=3.10$, the peaks corresponding to the self-distance of the two components do not overlap (cf.~Fig.~\ref{fig:asymmetry}c).
For $T>3.10$, the two self-distance peaks in $P(d_\text{avg})$ cannot be resolved at the level of the resolution provided by our numerical simulations.

In conclusion, in this section, we observe an alternative control problem that exhibits the topological transitions discussed in the main text, but does not possess the symmetry-breaking transition. 
Our result shows that, unlike the symmetry-breaking transition at $T_\text{sb}$, topological transitions at $T_\text{t--},T_\text{t+}$ do not depend on the symmetry $s(t) \leftrightarrow -s(T-t)$ of the control problem.

\end{document}